\documentclass[epj,final]{svjour}
\usepackage{graphicx}% \usepackage{dcolumn} \usepackage{amsmath} 
\usepackage[dvips]{color}
\usepackage{colordvi}
\usepackage{amsfonts}
\usepackage{amssymb}
\DeclareMathAccent{\pol}{\mathord}{letters}{"7E}
\newcommand{\he}{$^3$\rm{He}}
\newcommand{\hepol}{$\pol{^3\rm{He}}$}
\newcommand{\dpol}{$\pol{\rm{D}}$}
\newcommand{\heep}{$\pol{^3\rm{He}}$($\pol{e},e'p)$}
\newcommand{\heen}{$\pol{^3\rm{He}}$($\pol{e},e'n)$}
\newcommand{\hee}{$\pol{^3\rm{He}}$($\pol{e},e')$}
\newcommand{\hueep}{$\pol{^3\rm{He}}$(e,e'p)}
\newcommand{\hueen}{$\pol{^3\rm{He}}$(e,e'n)}

\newcommand{\BB}{$\pol{^3\rm{He}}$($\pol{e},e'p$)$d$}

\newcommand{\Apar}{$A_{\parallel}$}
\newcommand{\Aperp}{$A_{\perp}$}
\newcommand{\gen}{$G_{\rm{en}}$}
\newcommand{\gmn}{$G_{\rm{mn}}$}
\newcommand{\gep}{$G_{\rm{ep}}$}
\newcommand{\gmp}{$G_{\rm{mp}}$}
\newcommand{\emiss}{$E_{\rm m}$}

\newcommand{\et}{{\em et al.}}
\begin{document} 
\title{Experiments with polarized  $^{\bf 3}$He at MAMI}  
\author{A1 and A3 Collaboration \\\\
D.~Rohe
}

\institute{Departement f\"ur Physik und Astronomie, Universit\"at Basel, Klingelbergstr.82, 4056 Basel, Switzerland
}

\mail{D. Rohe, e-mail: Daniela.Rohe@unibas.ch}

\authorrunning{D. Rohe}
\titlerunning{Experiments with polarized  $^3$He at MAMI} 
\date{\today}

\abstract{Experiments with polarized \hepol\ at MAMI have already a long tradition. The A3 collaboration started in 1993 with the aim to measure the electric form factor of the neutron. At this time MAMI was the second accelerator where experiments with \hepol\ were possible. Some years before this pilot experiment the development of the apparatus to polarize \he\ in Mainz started. There are two techniques which allow to polarize sufficient large quantities of \he. Both techniques will be compared and the benefit of \hepol\ for nuclear physics will be discussed. An review of the experiments done so far with \hepol\ at MAMI will be given and the progress in the target development, the detector setup and the electron beam performance will be pointed out.    
 % end of PACS codes
}

\maketitle 

\section{Introduction}\label{intro}
Polarized \hepol\ has gained increasing interest due to its special spin structure described below, but also due to the fact that the Schr\"odinger equation for the three-body system can be exactly solved by means of the Faddeev formalism \cite{Golak02,Schulze93}. Further it is the only polarized target which tolerates currents of several $\mu A$ compared to $\approx$ 100~$nA$ for a N$\pol{\rm D}_3$ target. This helps to compensate the smaller thickness of the gas target. The gas target has the advantage that it is almost not diluted by unpolarized carrier material as it is the case for the N$\pol{\rm D}_3$ target.   

With the availability of highly polarized \he\ of several bars and the delivery of polarized continuous electron beams of high intensity, spin-dependent quantities can be studied, which show a large sensitivity to the underlying nuclear structure and reaction mechanism. Since in \he\ the protons reside with high probability in the $S$-state, the spin of \he\ is essentially carried by the neutron \cite{Blank84}. This property of the \he-spin structure can be best exploited in the quasielastic reaction \heen\ with restriction to small missing momenta as well as in inclusive \hee\ near the top of the quasielastic peak. In such kinematics the \hepol-target has been used extensively as polarized neutron target to measure the magnetic \cite{Gao94,Xu00,Xu03} and electric \cite{Meyerhoff94,Becker99,Golak01,Rohe99,Bermuth03} form factors of the neutron, \gmn\ and \gen.

Combining the theoretical calculation with the data gives insight into the three-body system and the nuclear structure of \hepol. Final state interactions (FSI) and Meson exchange currents (MEC) can be probed and studied under different kinematical conditions. There are also reactions and kinematics where \hepol\ does not appear as neutron target. In the \BB\ reaction, {\em e.g.}, \hepol\ appears as a polarized proton target \cite{Achen05}. Such a measurement will also be presented in this proceeding. 

The usual Faddeev calculations include FSI and MEC but are carried out non-relativistically. It was shown in \cite{Carasco03} that in particular relativistic kinematics plays an important role already at $Q^2$ = 0.67 (GeV/c)$^2$ (s. sec. \ref{sec_rel}). Less important are the relativistic treatment of the current operator and \he\ ground state. A relativistic ground state wave function became only recently available with the development of a Lorentz boosted nucleon-nucleon potential. It was constructed with the condition to give the same N-N phase shifts with the relativistic Lippmann-Schwinger equation as the non-relativistical potential when used with the Schr\"odinger equation  \cite{Kamada02}. The problem of the relativistic version of the Faddeev calculation is that it can treat only the interaction between the two nucleons which are not directly involved into the reaction (= spectators). We hope that further  ongoing theoretical work will lead to a full relativistic treatment of the three-body system. Experimental data will support such an effort.  

On occasion of the symposium this contribution aims to give a review of experiments performed with \hepol\ at MAMI in the last 20 years. The huge progress made during this time in the development of the target, the performance of the polarized electron beam and the improvement of the detector setup will be demonstrated. The different objectives for the experiments will be presented and finally the attempt is made to give an outlook on this field.

\section{Polarization methods}\label{polmeth}
For nuclear target applications two methods are in use, metastable-exchange optical pumping (MEOP) \cite{Walters62} and spin-exchange optical pumping (SEOP) \cite{Bouchiat60}. Both methods were already developed in the 60'ies but became only efficient in use with the development of laser light sources of sufficient power and proper frequency band width. MEOP is used for the Mainz target whereas at {\em e.g.} Jefferson Lab the SEOP techniques is applied. Both technique will be shortly explained and the advantages of each method discussed. It should be mentioned that there is a third method to polarize \he. Here high magnetic fields and low temperatures are needed which leads to polarizations of 38~\% in solid \he\ \cite{Haase98}. Due to the low heat conductivity of the solid \he\ this method is not suitable for nuclear physics experiments with electron beams.

In MEOP as well as in SEOP angular momentum is transferred to the atomic electrons by resonant absorption of circularly polarized light and subsequent reemission of unpolarized light. A magnetic field of 5 - 30 G defines the quantization axis. In MEOP an atomic transition in \he\ is pumped whose lower level is the metastable 2$^3$S$_1$ state. It is reached by a weak gas discharge (a fraction of 10$^{-6}$ atoms is excited). Therefore this method works only at low pressures of about 1 mbar which also guarantees a sufficiently long life time of the  2$^3$S$_1$ state. With moderate laser power of about 10 - 20 W and for large gas quantities of 20 liter at 1 mbar a polarization up to 80~\% can be reached in a minute. The formerly used LNA-Laser ($\lambda$ = 1083 nm, $\approx$ 10 W) is nowadays replaced by two Ytterbium fiber lasers (15 W each). Due to hyperfine coupling the electronic polarization results in a corresponding alignment of the nuclear spin. Subsequent collisions between polarized \he$^*$-atoms in the first excited metastable state and unpolarized \he-atoms in the ground state transfer the nuclear polarization to the ground state \he. The process of metastable-exchange collisions is fast and has a large cross section (10$^{-15}$ cm$^2$). Therefore this method is quite efficient.

In SEOP an alkali-vapour (usually Rb) is optically pumped by the circularly polarized light provided by a Ti-sapphire laser or by diode lasers tuned to the D$_1$ resonance line of 795 nm. Once the Rb is polarized, the polarization is transferred to the \he\ via spin-exchange collisions. The spin-exchange mechanism proceeds via the hyperfine interaction between the \he\ nucleus and the Rb valence electron. This can induce both species to flip their spin. Because this interaction is weak the cross section for this process is small (10$^{-24}$ cm$^2$). Therefore optically thick Rb vapour and large laser power ($>$ 40 W) are needed to polarize the gas in a target cell of 10 bar within 20 h to 50~\%. The advantage is that no further compression stage is required and a compact design is possible. To avoid radiation trapping in the optical thick Rb vapour which occurs when unpolarized resonant fluorescence light is emitted and afterwards reabsorbed, 50 - 100 mbar nitrogen is added. The addition of a fraction of 10$^{-2}$ N$_2$ leads to $\approx$ 5 (10) ~\% contribution to the scattering rate from a proton (neutron) and therefore to a effective dilution of the polarization observables.  

Except for experiments in a storage ring the mass density of a few mbar of polarized \hepol\ from MEOP is too low for a nuclear physics experiment. Therefore one or two mechanical compression stages\footnote{In the Hermes experiment the cell was cooled down to 25 K to achieve a compression factor of 3.5.} are necessary to reach pressures of up to 6 bars. Up to now three different polarizers were in use for nuclear physics experiments at MAMI. The first one was the Toepler compressor which uses 17.6 kg mercury as a piston. The pressure achieved in the 100 cm$^3$ target cell was 1 bar and the polarization could be increased from 38~\% to 49~\% from 1993 to 1995. The target cell was polarized in a continuous flow (0.1 bar l/h) and the polarization loss from the low pressure pumping cell to the target was 30~\%. The increase of the polarization was achieved by coating the target cell with cesium to reduce the relaxation of the polarization due to collisions with the container material (glass). The Toepler compressor was developed for the first measurement of the electric form factor of the neutron \gen\ which is described below. Nowadays the compression stage is replaced by one titanium piston which allows a production rate of 1.5 bar l/h. The polarization losses are negligible and the target cell is filled with 5 bar and 75 \% polarization. This is a great improvement and increases the performance of the nuclear physics experiment significantly. 

\section{Experiments with polarized $^{\bf 3}$He}
\subsection{The electric form factor of the neutron}\label{ffel}
\subsubsection{Motivation and techniques}\label{ff_mot}
Form factors describe the response of the nucleon to the interaction. For spin 1/2 particles there are two form factors determining the electromagnetic response, the magnetic and electric form factor. They are related via a Fourier transformation to the magnetic and charge distribution (s. sec. \ref{charge}), respectively. A form factor independent from the momentum transfer to the particle would indicate a pointlike distribution. Any deviation from it points to an underlying substructure. The electric form factor \gen\ of the neutron is particularly sensitive to its internal structure because it is not obscured by the total charge as in the proton. The substructure of the nucleon is determined by the (sea-)quarks and the gluons. Therefore \gen\ is a good test case for our understanding of the quark degrees of freedom and a constraint for models. QCD would be the first choice to calculate form factors but it is still limited due to the computer power available. Often approximations (quenched lattice) are applied to avoid the computationally expensive part. Recently a full lattice QCD could reproduce the trend of the data \cite{Goeck05}. With the extension of the data base in the last few years the theoretical interest also increases and a large variety of models and model-based fits were developed. This was not the case in 1987 when the first \gen\ measurement at MAMI was planned. The data base was scarce and in particular the error bars exceed 100~\%. The reason: \gen\ is difficult to measure, as its value is small, roughly a factor 10 smaller than the magnetic form factor \gmn. The nucleon form factors enter the quasielastic cross section quadratically, so the magnetic scattering amplitude dominates by far. Therefore a LT separation in the reaction (e,e'n) leads to unreasonably large errors. A further complication comes from the fact that there exists no free neutron target of sufficient density. Thus the deuteron or \he\ have been employed. This leads to corrections due to the nuclear structure and to contributions from the much larger form factors of the proton.

\begin{figure}[t]
\begin{center}
\includegraphics[width=8.7cm,clip]{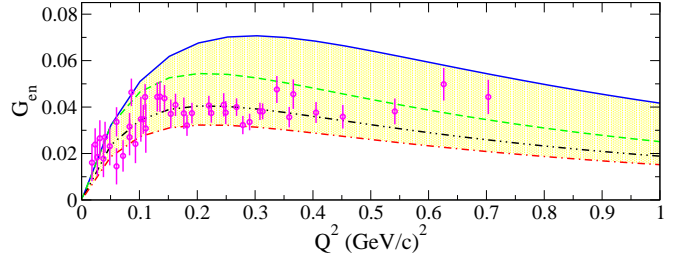}
\end{center}
\caption{\label{platch}The data points have been determined from elastic electron-deuteron scattering using the Paris potential. The curves are fits to the data using different N-N potentials (Nijmegen, Argonne V14, Paris, Reid Soft Core from top to bottom) and demonstrate the model dependence of the resulting \gen.}
\end{figure}

In 1990 the best data were measured by Platchkov and collaborators \cite{Platch90} using elastic scattering on the deuteron. A LT separation gives the longitudinal and the transverse structure functions A(Q$^2$) and B(Q$^2$). A(Q$^2$) depends quadratically on the charge and quadru\-pole form factors of the deuteron. The charge form factor dominates for small Q$^2$ ($<$ 0.4 (GeV/c)$^2$). It is proportional to (\gep\ + \gen)$^2$ and therefore contains an interference term \gep\ times \gen\ which increases the sensitivity to \gen. On the other hand the contribution from \gep$^2$ had to be removed which increases the uncertainty in \gen. The main drawback is that A(Q$^2$) contains also the deuteron structure which depends on the nucleon-nucleon potential chosen to calculate the deuteron form factor. This introduces a large model dependence of about 50~\% shown in fig. \ref{platch} \cite{Platch90}. The choice of a modern N-N potential would lead to a smaller model uncertainty. The treatment of MEC, which makes a significant correction, introduces further uncertainties. 

The systematic errors described above can be significantly reduced by exploiting the quadrupole form factor of the deuteron instead of $A(Q^2$). The contribution from two-body currents is relatively small and the sensitivity of \gen\ to the choosen N-N potential is reduced \cite{Schia01}. However, at low $Q^2$ the statistical error of $F_{C2}$ is large because the monopole form factor $F_{C0}$ dominates the  $T_{20}$ data. Thus, the analysis using $A(Q^2)$ becomes superior for $Q^2$ $<$ 0.4 (GeV/c)$^2$.  

A method which is much less model dependent exploits the observables measurable in a double-polarization experiment. In exclusive reactions it is a sensitive tool to measure \gen. Here the longitudinally polarized electron beam scatters quasielastically on deuterons or $^3$He, which are either polarized or where the polarization of the knocked-out neutron is detected \cite{Arnold81} (s. contribution of M. Ostrick to this symposium). The asymmetry with respect to the electron helicity contains then an interference term \gen\ times \gmn\ which amplifies \gen\ by \gmn. The sensitivity to \gen\ is largest in the perpendicular asymmetry \Aperp, where the direction of the target spin is perpendicular to the momentum transfer (or the polarization of the scattered neutron is perpendicular to its momentum, respectively). In contrast the parallel asymmetry \Apar\ does not depend on form factors (for \gen\ small) and therefore can serve as normalization. Measuring the asymmetry has the advantage that no absolute cross section measurements are required which avoids the effort (and systematic errors) of determining absolute efficiencies, solid angles and luminosity. 

The electron-target asymmetry is obtained via
\begin{equation} \label{asymexp}
A_{exp} = \frac{N^+/L^+ - N^-/L^-}{N^+/L^+ + N^-/L^-},
\end{equation}
where $L^+$ ($L^-$) are the integrated charge and  $N^+$ ($N^-$) the number of events for positive (negative) electron helicity. The electron helicity is flipped every second randomly.

In general, the asymmetry $A$ can be decomposed according to the direction of the target spin which is given by the angles $\theta_S$ and $\phi_S$  with respect to the momentum transfer $\pol{q}$ and the scattering plane. 
\begin{equation} \label{asymtheo}
A = A_{\perp} \sin{\theta_S} \cos{\phi_S} + A_{\parallel} \cos{\theta_S}
\end{equation}
Before the asymmetry obtained in the experiment can be compared to theory it has to be corrected for the polarization of the electron beam $P_e$ and the target $P_T$ as well as for a dilution factor V.
\begin{equation}
A = \frac{1}{P_e P_T V} A_{exp}. 
\end{equation}
The dilution factor $V$ can come from the scattering on unpolarized carrier material in the target or scattering on the target container (background). Also charge exchange p $\rightarrow$ n in the shielding in front of the hadron detector contributes to $V$ because the protons in \hepol\ are almost unpolarized. The corrected asymmetry $A$ contains the electromagnetic form factors but also depends on the reaction mechanism involved (s. below). For scattering on a free neutron with polarization $P_n$ one has
\begin{eqnarray} 
A_{\perp} &=& \frac{1}{P_e P_n} \frac{2\sqrt{\tau(1+\tau)} \tan(\theta/2) G_{en} G_{mn}}{G^{2}_{en}+G^{2}_{mn} (\tau + 2 \tau (1 + \tau) \tan(\theta/2))} \label{asymfree1}\\ \vspace*{5mm}
A_{\parallel} &=& \frac{1}{P_e P_n} 2 \frac{\tau \sqrt{1 + \tau + (1+\tau)^2 \tan^2(\theta/2)} \tan(\theta/2) G^{2}_{mn}}{G^{2}_{en}+G^{2}_{mn} (\tau + 2 \tau (1 + \tau)) \tan(\theta/2)}\nonumber. \label{asymfree2}
\end{eqnarray}
\gen\ is determined best from the ratio of the asymmetries \Aperp\ and \Apar\
\begin{equation} \label{asymratio}
\frac{A_{\perp}}{A_{\parallel}} \varpropto \frac{G_{en}}{G_{mn}}
\end{equation}
instead of \Aperp\ alone. This has several advantages. The polarization product $P_e$ $P_T$ drops out. Therefore the systematic error introduced with the two measurements of absolute polarizations can be considerably reduced. In addition, theoretical corrections accounting for the nuclear structure in \hepol\ are reduced in the ratio. The polarization of the neutron $P_n$ bound in \hepol\ and entering eq. \ref{asymfree1}, e.g., is usually smaller than the polarization $P_T$ of \hepol\ measured. Further the dilution factor $V$ cancels.
 
\subsubsection{Form factor measurements in A3}\label{ff_A3}
\begin{figure}[t]
\begin{center}
%%36 40 550 776
\includegraphics[width=6cm,clip]{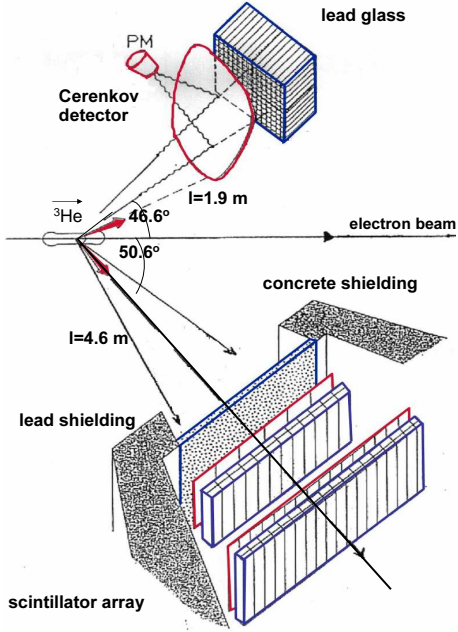}
\end{center}
\caption{\label{A3setup}Setup of the \gen\ experiment in the A3 hall at MAMI. Electrons are detected in the segmented lead glass detector in coincidence with neutrons in the plastic scintillator array.}
\end{figure}
The setup for the \gen\ experiment in the A3 experimental hall is shown in Fig. \ref{A3setup}. Electrons were detected in a calorimeter of 256 closely packed lead glass counters (4 x 4 cm) with a solid angle of 100 msr in a distance of 1.9 m. The shower produced by the electron extends over about 10 modules. Therefore the energy summed over clusters of detectors was used leading to an energy resolution $\Delta$E/E = 20~\% FWHM. This moderate energy resolution was sufficient to separate the inelastic contribution from the quasielastically scattered events. The inelastic events, mainly resulting from $\pi$-production in the $\Delta$-resonance, have vanishing asymmetry. In case of an admixture this would dilute the asymmetry. In front of the calorimeter a focusing air Cerenkov detector was placed which suppresses background from electrons scattered on the exit or entrance windows of the target cell or the beam line. Further it serves to discriminate photons and pions from electrons.

The neutrons were detected in a plastic scintillator array which covers 250 msr and therefore the entire Fermi cone. It consisted of two walls and could also be used as a neutron polarimeter for the \gen-measurement using the reaction D($\pol{e}$,e'$\pol{n}$) \cite{Herberg99,Ostrick99}. The overall detector thickness of 40 cm yields a neutron detection efficiency of $\epsilon_n$ = 32~\%. The neutron detector was shielded with 5 cm lead on the front and surrounded by 1 m concrete against electromagnetic background.

In this setup the already mentioned Toepler compressor was used to produce 1 bar \hepol\ with polarizations increasing from 38~\% to 49~\% from 1993 to 1995. At the same time the electron polarization could be increased from 30 to 50~\% by changing the cathode from a bulk to a strained layer GaAsP.  Keeping in mind that the statistical error of the asymmetry decreases with (P$_e$ P$_T$ $\sqrt{T}$)$^{-1}$ (T: measurement time) both improvements enhance the performance of the experiment significantly.

\begin{figure}[t]
\begin{center}
\includegraphics[width=8cm,clip]{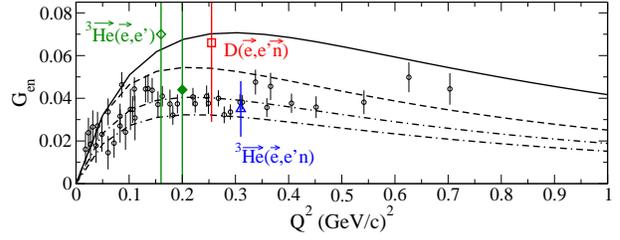}
\end{center}
\caption{\label{firstresults} (Uncorrected) result from the first measurement of \gen\ at MAMI using \hepol\ \cite{Meyerhoff94}. Two double-polarization experiments performed at Bates in the same time period are also shown \cite{Thompson92,Eden94}. For comparison the data already shown in fig. \ref{platch} are displayed as well.}
\end{figure}

With this setup \gen\ was measured at Q$^2$ = 0.35 (GeV/c)$^2$ \cite{Meyerhoff94,Becker99}. The pilot experiment of Meyerhoff in 1993 \cite{Meyerhoff94} did only use a quarter of the detector setup shown in fig. \ref{A3setup}. Its result is shown in fig. \ref{firstresults} together with other double-polarization experiments performed at Bates at the same time using $\pol{\rm ^3He}$($\pol{e},e')$ \cite{Thompson92} and  D($\pol{e},e'\pol{n}$) \cite{Eden94}. In fig. \ref{firstresults} the uncorrected results are shown. In the mid of the 90'ies it was not clear that the measured value for \gen\  using \hepol\ as polarized neutron target needs a large correction accounting for FSI. No exact Faddeev calculation was available and the diagrammatic approach of Laget \cite{Laget91,Laget92} indicated a negligible correction. Later is was shown by the Bochum-Krakow group that a correction of about 30~\% had to be applied on \gen\ at Q$^2$ = 0.35 (GeV/c)$^2$ \cite{Golak01}. 

\subsubsection{Non-PWIA contributions}\label{Ay}
An experimental measure for non-PWIA contributions is the target analyzing power A$_y$. For coplanar scattering  A$_y$ is identical to zero in PWIA due 
to the combination of time reversal invariance and hermiticity of the transition matrix \cite{Conzett98}. Thus, a non--zero value of A$_y$ signals FSI and MEC effects and its measurement provides a sensitive check of the calculation  of these effects. For an unpolarized beam and the target spin aligned perpendicular to the scattering plane the target analyzing power can be measured
\begin{equation} \label{eq_Ay}
A_{y}^o = \frac{1}{P_T} \frac{N^{\uparrow} - N^{\downarrow}}{N^{\uparrow} + N^{\downarrow}},
\end{equation}
where $N^{\uparrow},(N^{\downarrow})$ are the normalized 
$\mathrm{^3\overrightarrow{\mathrm{He}}(e,e'N)}$ events for target spin 
aligned parallel (antiparallel) to the normal of the scattering plane. This quantity was measured for the reactions \hueep\ and \hueen\ at Q$^2$  = 0.37 and 0.67 (GeV/c)$^2$ \cite{Bermuth03} using the experimental setup in the A1 spectrometer hall at MAMI described below. For this the target spin, aligned perpendicular to the scattering plane, was reversed every 2 minutes. Contrary to the determination of \gen, dilution effects do not cancel for A$_y$ in eq. \ref{eq_Ay} and have to be determined. The main contribution for \hueen\ comes from charge exchange in the 2 cm lead shielding in front of the hadron detector. This factor was determined using hydrogen as target. Then the recoil proton was tagged with the elastically scattered electrons in the spectrometer and the number of neutrons detected in the scintillator were counted. Due to the large difference in the cross section for (e,e'p) and (e,e'n) as well as in the detection efficiency the contribution from charge exchange to \hueep\ is negligible whereas for \hueen\ the correction is in the order of 10 to 15~\%. 

\begin{figure}[t]
\begin{center}
\includegraphics[width=8.7cm,clip]{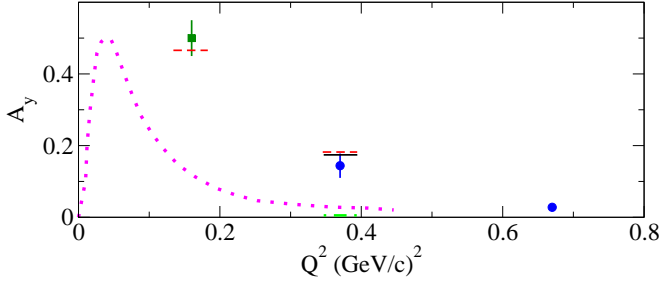}
\end{center}
\caption{\label{laget_Ay} Target asymmetry in the reaction ${^3\pol{\rm He}(e,e'n)}$ as a function of Q$^2$ measured at MAMI \cite{Bermuth03} (circles) and at NIKHEF \cite{Poolman99} (square). Shown is the result of the non-relativistic Faddeev calculation including FSI and MEC (solid) and FSI only (dashed). The dot-dashed line is obtained when neglecting charge-exchange (\gep\ = \gen\ = 0). The dotted curve represents the result from a diagrammatic approach \cite{Laget91}.}
\end{figure}

The corrected experimental result \cite{Bermuth03} for the reaction \hueen\ is shown in fig. \ref{laget_Ay} together with the data point measured at NIKHEF \cite{Poolman99}. Further the calculation from the Bochum-Krakow group is shown including FSI (dashed) and FSI plus MEC contributions (solid). The effect from MEC is small as expected for quasifree kinematics. A more detailed examination revealed  that the large contribution from FSI at small Q$^2$ comes mainly from (e,e'p) followed by charge exchange. The full calculation is in good agreement with the data. Also shown is a early result from Laget \cite{Laget91}. Clearly it underestimates the effect from non-PWIA reaction mechanism. Fig. \ref{laget_Ay} confirms that the FSI contribution and thus the theoretical correction to \gen\ gets smaller with increasing Q$^2$. This is expected from simple arguments like the decreasing of charge-exchange cross section with Q$^2$ and the shortening of the interaction time during the reaction at higher momentum transfer. 

\subsubsection{Form factor measurements in A1}\label{ff_A1}
To avoid a large theoretical correction the \gen\ measurement was extended to higher Q$^2$. A pilot experiment was performed already in 1997 in the A1 spectrometer hall at a Q$^2$ of 0.67 (GeV/c)$^2$ followed by a second experiment in 2000 to double the statistics. Using the same setup data on A$_y$ were taken (s. sec. \ref{Ay}). Both the target and detector setup were considerably improved compared to the experiment in the A3 hall. 

\begin{figure}[t]
\begin{center}
\includegraphics[width=8cm,clip]{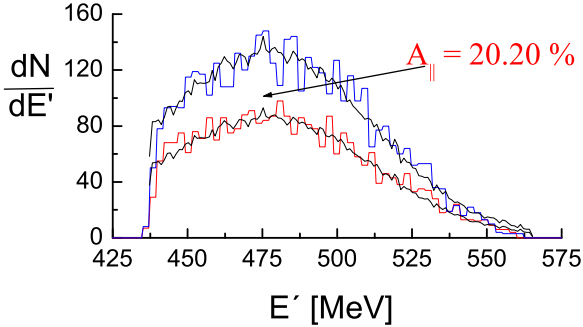}
\end{center}
\caption{\label{spec_mom} Electron momentum spectrum measured in spectrometer A for two electron helicities. The solid line corresponds to a Monte Carlo simulation.}
\end{figure}

The scattered electrons were detected in the magnetic spectrometer A which has a focal plane detector consisting of two drift chambers, a scintillator array and a Cerenkov detector. It has a momentum acceptance of 20~\% and a solid angle of 28 msr. Due to its high resolution and the efficient pion rejection in the Cerenkov the inelastic contribution can be well separated from the quasielastic region. Further the spectrometer serves to determine the direction of the momentum transfer $\pol{q}$ with good precision. The angle between $\pol{q}$ and the target spin direction has to be precisely known for the extraction of \gen\ from \Aperp. A momentum spectrum of the scattered electrons is shown in fig. \ref{spec_mom}. The two spectra belong to different electron helicities and target spin parallel to the momentum transfer. The thick solid line represents a Monte Carlo simulation using simple kinematical relations valid in Born approximation. The good agreement supports that the contribution from non-quasielastic events is negligible. 

The hadron detector was placed in direction of $\pol{q}$. It consists of four layers with five scintillator bars each. In front of the detector two layers of $\Delta$E detectors discriminate protons and neutrons. In 160 cm distance from the target the detector has a solid angle of about 100 msr. The entire detector was shielded with 10 cm lead except for an opening towards the target where a reduced shield of 2 cm was used. In addition a lead collimator in front of the detector helps to suppress background produced in the downstream beam line.   

The entire \hepol--target was enclosed in a rectangular box of 2~mm thick $\mu$-metal and iron except for a cut-out towards the opening angle of the spectrometer. The box served as an effective shield for the stray field of the  magnetic  spectrometers and provided a homogeneous magnetic guiding field of $\approx 4 \cdot 10^{-4}$~T produced by three independent pairs of coils. With additional correction coils a relative field gradient of less than $5 \cdot 10^{-4}$\,cm$^{-1}$ was achieved. The setup also allowed for an independent rotation  by remote control of the target spin in any desired direction with an accuracy of 0.1$^o$. 

The polarization of \he\ was monitored with Adiabatic Fast Passage (AFP) using the technique described in ref. \cite{Wilms97} which measures the magnetic field of the oriented spins. Since the AFP-technique destroys part of the polarization ($\approx$ 0.1 -- 0.2~\%) and since it cannot be used during data-taking due to spin-flipping, it is used only about once in 4 h. Therefore Nuclear Magnetic Resonance (NMR) monitored continuously ($\approx$ every 10 min) the relative polarization and served mainly as online control of the polarization. The systematic error of the absolute polarization is estimated to be 4~\% and the uncertainty in the relaxation time is 2~h. In the first (second) beam time in 1997 (2000) the averaged target polarization was 32~\% (36~\%).     

The \hepol--target consisted of a spherical glass container (diameter 9 cm) with two cylindrical extensions sealed with oxygen--free 25$\mu$m Cu--windows. The Cu--windows were positioned outside of the acceptance of the spectrometer ($\sim$5~cm) and shielded with Pb--blocks to minimize background from beam--window interactions. The $^3{\rm He}$--target was polarized via metastable optical pumping to a typical polarization of 0.5 and compressed to an operating pressure of 5~bar  with a two-stage titanium compressor \cite{Surkau97}. Then the target cell was transported in a portable magnetic field to the target pivot. The relaxation time of the polarization due to contact with the surface is increased by careful cleaning and coating with cesium to 80 h. The relaxation time is reduced to about 40 h due to the dipole-dipole interaction between the \he-atoms at high pressure and due to ionization of \hepol\ by the electron beam. This leads to the creation of $^3$He$_{2}^{+}$ and loss of polarization by transfer of angular momentum to the rotational degrees of freedom. The electron current used was 10~$\mu$A and the polarization 75 - 80~\% which was measured with a Moeller polarimeter installed a few meters upstream of the target pivot in the A1 three-spectrometer hall. Nowadays currents of more than 20~$\mu$A are routinely provided.

\begin{figure}[t]
\begin{center}
\includegraphics[width=8cm,clip]{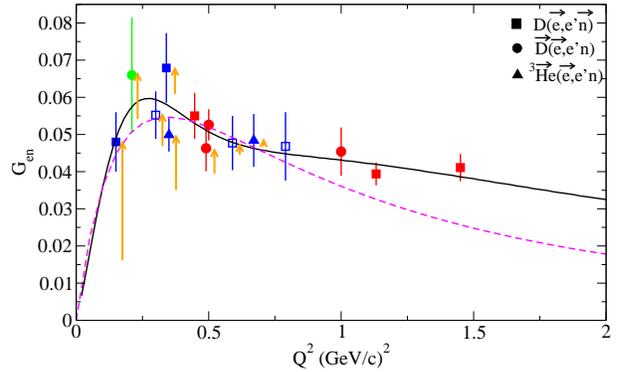}
\end{center}
\caption{\label{all_gen}\gen\ extracted from quasielastic scattering of polarized electrons from D, \dpol\ and \hepol. The data are taken from refs. \protect{ 
\cite{Passchier99,Herberg99,Ostrick99,Becker99,Zhu01,Bermuth03,Madey03,Warren04,Glazier05}}. The dashed line represents the Galster fit \cite{Galster71} and the solid curve the result of  \cite{Fried03}. For some of the experimental data points the correction due to the reaction mechanism beyond PWIA is indicated by the size of the arrows.}
\end{figure}  

Combining the measurement of A$_y$ (sec. \ref{Ay}) and the Faddeev calculation of the Bochum-Krakow group one estimates a correction to \gen\ at Q$^2$ of 0.67 (GeV/c)$^2$ of (3.4 $\pm$ 1.7) \%. At this Q$^2$ a relativistic calculation is already needed (s. sec. \ref{sec_rel}). The corrected \gen-value is shown in fig. \ref{all_gen} with all published results from double-polarization experiments. These measurements were performed at NIKHEF, Jlab and MAMI using polarized deuterium instead of \hepol\ or detecting the polarization of the knocked-out neutron in a polarimeter. In fig. \ref{all_gen} also indicated are the theoretical corrections applied to the \gen-value extracted from the data. Clearly the correction decreases with increasing Q$^2$. 

The dashed line in fig. \ref{all_gen} is a fit to data available in the 70'ies. Most of these data were obtained using elastic electron-deuteron scattering. As mentioned in sec. \ref{ff_mot} this method implies a large model dependence. The so called Galster fit was obtained by using data up to $Q^2$ = 0.8 (GeV/c)$^2$ with large statistical uncertainty \cite{Galster71}. For this fit the dipole form is modified in such a way that the slope at small $Q^2$ known from n-e scattering could be reproduced. Surprisingly this fit still gives a good description of the actual data set up to $Q^2$ = 0.8 (GeV/c)$^2$. 

%\subsubsection{Theoretical description of form factors}\label{theo_ff}

\subsubsection{Charge distribution of the neutron} \label{charge}
A new fit to the present data set was provided by Friedrich and Walcher \cite{Fried03} using a phenomenological model of the nucleon. In this model a superposition of two dipoles for the smooth part and two Gaussians to account for a possible bump is used as fitting function. Their result is shown by the solid line in fig. \ref{all_gen}. Here a bump at Q$^2$ $\approx$ 0.2 (GeV/c)$^2$ describes the data set best but more precise data are needed for a firm confirmation. It is remarkable that also in \gmn, \gep\ and \gmp\ this bump appears in the same Q$^2$ region. Due to the large magnetic moments of the nucleons and the charge of the proton it is only visible if the form factor data are divided by the dipole form factor.  

A physically motivated fit decomposes the neutron into a bare neutron n$_o$ and a polarization part: 
\begin{equation} \label{neutron}
n = (1 - b_n) n_o + b_n (p_o + \pi^-)
\end{equation}
The bare neutron consists of three quarks with a form factor assumed to be of  the dipole form. In the polarization part the neutron exists as a bare proton p$_o$ surrounded by a pion cloud. The form factor of the pion is constructed from the spatial distribution of the harmonic oscillator wave function in a p-state. With six free parameters a good description of both \gen\ and \gep\ is achieved. According to this fit the neutron exists to 90~\% as bare neutron and to b$_n$ $\approx$ 10~\% as proton with pion cloud. 

\begin{figure}[t]
\begin{center}
\includegraphics[width=8cm,clip]{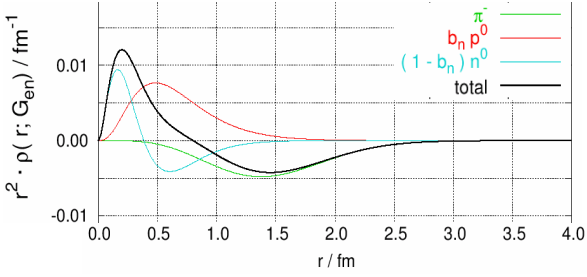}
\end{center}
\caption{\label{rho_distr}Charge distribution (weighted with the radius squared) of the neutron decomposed into the contributions from the bare proton p$_o$, the pion cloud and the bare neutron n$_o$ (picture taken from \cite{Fried03}).}
\end{figure}  

From such a fit the charge distribution of the neutron can be obtained by a Fourier transformation,
\begin{equation}
\rho(r) = \frac{1}{2 \pi^2} \int_{0}^{\infty} G_{en}(Q^2) \frac{\sin(Qr)}{Qr} Q^2 dQ
\end{equation}
where Q is the momentum transfer in the Breit frame, i.e. a frame with no energy transfer to the nucleon. The behaviour at high Q$^2$ mainly determines the charge distribution deep inside the nucleon at small radii r. In fig. \ref{rho_distr} the result for the charge distributions  $\rho$(r) r$^2$ of the neutron (black line) as well as for the three components in eq. \ref{neutron} are shown. It is remarkable that the pion contribution extends as far as 2 fm (maximum at $\approx$ 1.5 fm).  In contrast, the authors of ref. \cite{Ham04} separated the contribution of the two-pion continuum and found a peak at a distance of only 0.3 fm. The maximum of the pion cloud in the model of Friedrich and Walcher corresponds to the Compton wavelength of the pion ($\lambda$ = 1.43 fm) determining the range of the nuclear force in the Yukawa model. This confirms that in this model only one pion is taken into account by construction. This consideration might resolve part of the disagreement because in ref. \cite{Ham04} two-pion contributions are considered. 

\subsection{Test of the theory at Q$^2$ = 0.67 (GeV/c)$^2$} \label{sec_rel}
The Faddeev calculation mentioned so far is fully non-relativistic and it was not clear at which Q$^2$ relativistic effects would become non-negligible. On the other hand the contribution from non-PWIA reaction mechanisms to the asymmetries in \heen\ are small at high Q$^2$ as shown in sec. \ref{ff_A1}. More sensitive are the asymmetries  \Apar\ and \Aperp\ in the reaction \heep. These asymmetries are small because the two protons are most of the time in the S-state. In this case the asymmetries vanish. Therefore comparing the experimental result to the theory provides a sensitive test to effects from reaction mechanism as well as from the \hepol\ structure. Both might be influenced by a relativistic treatment.

There are several ingredients in the Faddeev calculation which might be treated relativistically or non-rela\-ti\-vistically. This includes the 1-body current operator, the T-matrix element describing the FSI, the kinematics and the \hepol\ ground state wave function. It should be mentioned that up to now in the relativistic description only the interaction between the spectator nucleons, i.e. the ones which are not involved in the primary reaction, can be included. This is called FSI23 or rescattering term of first order. At the moment there are no exact calculations available for \he\ which can treat MEC and full FSI at high $Q^2$. The relativistic treatment of the \hepol\ ground state became only recently available with the development of a Lorentz boosted NN potential. In ref. \cite{Kamada02} such a potential was obtained and used in a relativistic 3N-Faddeev equation for the bound state to calculate the triton binding energy. The results presented below are still based on an exact but non-relativistic \he\ ground state. A newer calculation prepared for a recent proposal to measure \gen\ at Q$^2$ of 1.5 (GeV/c)$^2$ shows that the difference is small.  

The dependence on the NN--interaction was studied with a calculation which employs the CD-Bonn NN-poten\-tial \cite{Machleidt96} instead of the AV18 NN-potential \cite{Wiringa95}. The difference in the result is negligible. It should be mentioned that the potential approach is not strictly valid when the center-of-mass energy of the three-nucleon (3N) system is well above the pion production threshold. The c.m. energy available in the 3N-system $E_{3N}$ can be obtained for the 3-body breakup via
\begin{equation}
E_{3N} = \sqrt{(M_{He} + \omega)^2 - |\vec{q}|^2} - 2 M_p - M_n
\end{equation}  
However, in quasi--elastic kinematics the focus is mostly on the region of phase space, where one of the nucleons is struck with a high energy and momentum and leaves the remaining two-nucleon system with a rather small internal energy. Thus this  approximations which has to be made also in other calculations for \he\ and deuterium \cite{Deltuva04,Aren92}, might not too seriously influence the result. 

\begin{figure*}[hbt]
\begin{center}
\includegraphics[scale=0.4,clip]{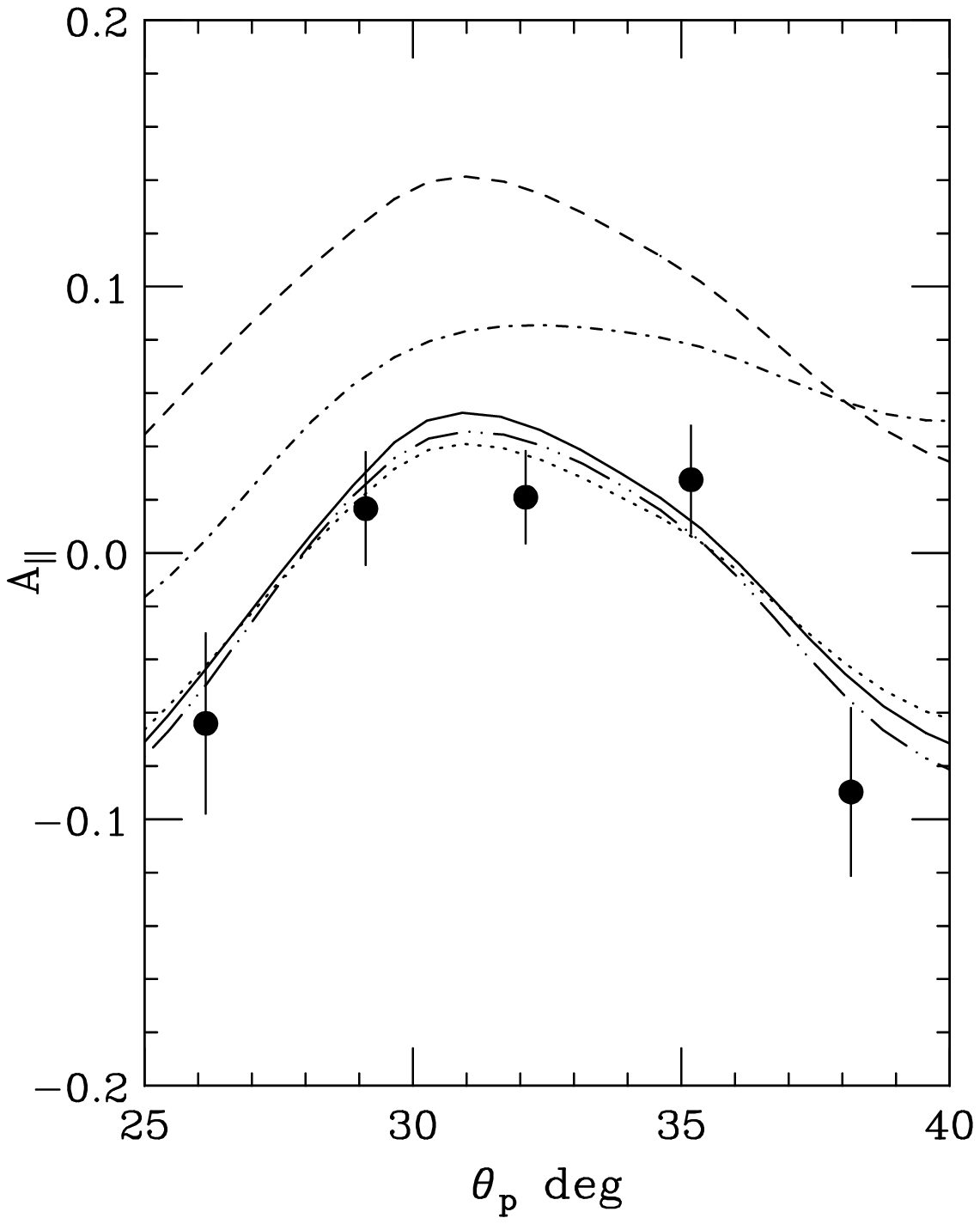}
\hspace*{1.5cm}
\includegraphics[scale=0.4,clip]{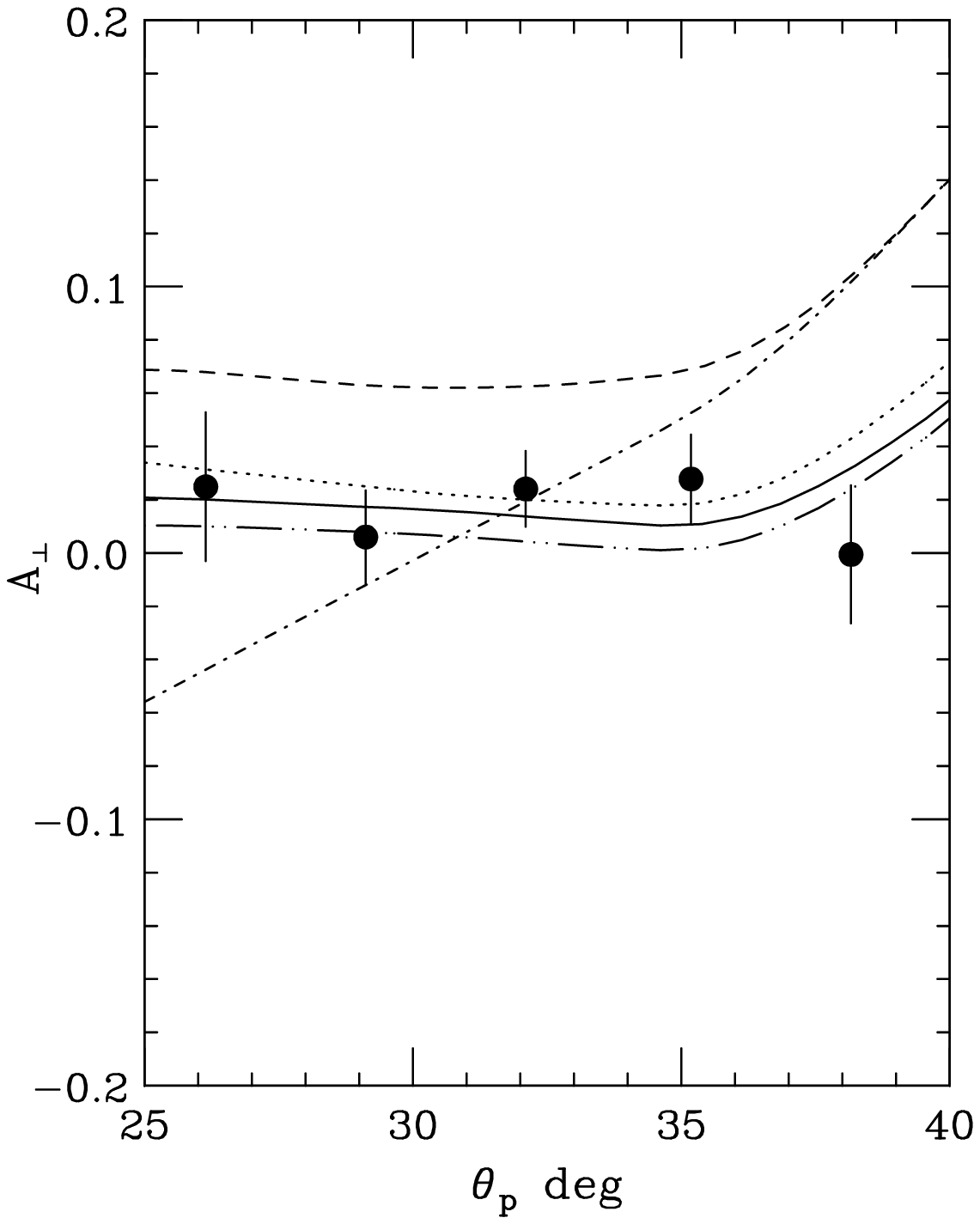}
\parbox{13cm}{\caption[]{\label{eep}
Experimental results of \Apar\ (left) and \Aperp\ (right) for the central region of the quasi--elastic peak as a function of the scattering angle of the knocked-out proton. The results of the full (PWIA) calculation are shown with solid (dashed) lines. The result of the full calculation with a non--relativistic current (dot), the effect of a (v/c)$^2$ correction (dot-dot-dash) and the same with non--relativistic kinematics (dot-dash) are also shown.
}} 
\end{center} 
\end{figure*}

The data on the reaction \heep\ were taken simultaneously to the measurement of \gen\ at Q$^2$  = 0.67 (GeV/c)$^2$. Protons were selected in the hadron detector by requiring hits in two consecutive $\Delta$E detectors. The background in the coincidence time spectrum, determined from the time difference between the first bar in the hadron detector and the scintillator plane of spectrometer A, was negligible. 
In order to study the effect of FSI on the asymmetries in different kinematic regions, the quasi--elastic peak is divided into two regions of $\omega$. One region covers the peak and 
therefore emphasizes low nucleon momenta whereas the other region covers the low $\omega$ tail sensitive preferentially to high nucleon momenta. The events in each of the two regions are summed over the entire acceptance of the out--of--plane angle of electron and proton and over the electron scattering angle in a range from 75.8$^o$ to 81.8$^o$. 

In Fig. \ref{eep} the parallel and perpendicular asymmetries in the central region of the quasielastic peak are shown as a function of the scattering angle of the proton. They are compared to the theory which contains the two-body (2BB) and three-body break\-up (3BB). The 3BB channel is integrated over the first 26 MeV. As can be seen from the figures the PWIA calculation (dashed line) clearly disagrees with the data. From the calculations which include FSI23 only, the one with non-relativistic kinematics (dot-dashed line) cannot describe the experimental results. Relativistic (solid line) or non-relativistic (dots) treatment of the current operator does not make a large difference. The calculation taking into account relativistic kinematics and FSI23 provides a good description of the data. Both ingredients are important to achieve agreement with the experimental results. 

\subsection{Structure of $^{\bf 3}\pol{\bf \rm He}$} \label{struc}
In the experiment described in the previous section it was not possible to separate the 2BB and 3BB channel due to the limited resolution of the hadron detector. For a better understanding of the structure of \hepol\ spectrometer B was taken for proton detection. The kinematics was limited to the central region of the quasielastic momentum distribution at Q$^2$ of 0.31 (GeV/c)$^2$. Each hour the target spin was turned to measure the parallel, perpendicular, antiparallel and antiperpendicular asymmetry alternately with the purpose to reduce the systematic errors. The target cell was of the same kind as already used for the \gen\ measurement (s. sec. \ref{ff_A1}). A new polarizer was used consisting of one-stage titanium compressor \cite{Otten04} with effectively no polarization loss during the transfer from the low pressure gas reservoir to the target container. The \he\ was optically pumped with two Ytterbium fiber lasers each providing 15~W on the resonance transition (1083 nm). With this setup an initial target polarization of 70 to 75~\% could be achieved.  Averaged over the beam time period and accounting for relaxation a target polarization $P_{\rm T}$ of (49.8  $\pm$ 0.3 (stat.) $\pm$ 2 (syst.))\% was obtained. 

\begin{figure}[t]
\begin{center}
\includegraphics[width=7cm,clip]{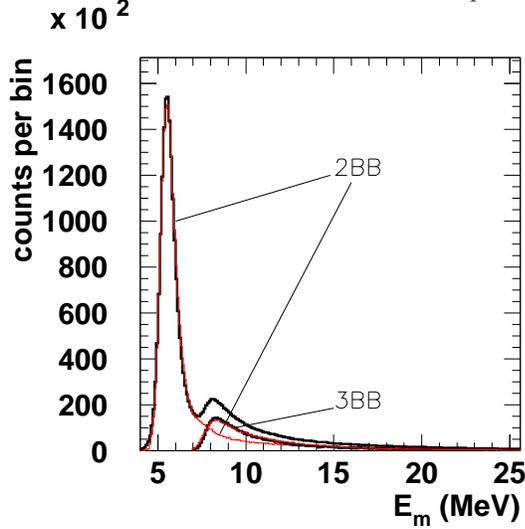}
\end{center}
\caption{\label{Emsimulin}Experimental \emiss\ distribution (thick line 2BB) and the
simulation of the 2BB (thin red line). The difference is shown as thick black line 3BB.}
\end{figure}

From the measured kinematic variables in the two spectrometers, the missing energy is reconstructed according to
\begin{equation} \label{Em}
E_m = E - E_e - T_p - T_R .
\end{equation}
Here, $E$ ($E_e$) is the initial (final) electron energy and $T_p$ is the kinetic energy of the outgoing proton.  $T_R$ is the kinetic energy of the (undetected) recoiling (A--1)-system, which is reconstructed from the missing momentum under the assumption of 2BB. The resulting \emiss\ distribution reconstructed from the data is shown in fig. \ref{Emsimulin} as thick solid line. The resolution is limited mainly by the properties of the target cell and not by the resolution of the spectrometers. The FWHM of 1~MeV allows a clear separation of the \emiss-regions where only 2BB or 2BB and 3BB contribute. The \emiss-region from 4.0 to 6.5 MeV is interpreted as pure 2BB. This cut was chosen to avoid any contribution from the 3BB-channel (starting at 7.7 MeV) considering the experimental \emiss\ resolution. In agreement with ref. \cite{Flori99}, the yield of the 3BB is negligible beyond 25 MeV. Therefore the cut for the 3BB-channel was made from 7.5 to 25.5 MeV in the \emiss\ spectrum. Because the 3BB resides on the radiation tail of the 2BB, the latter has to be accounted for in the analysis of the 3BB-region of the measured spectrum. To this end, the tail was calculated in a Monte Carlo simulation which accounts for internal and external bremsstrahlung, ionization loss and experimental energy resolution adjusted to the experimental distribution. The simulated 2BB distribution is shown as thin red line in fig. \ref{Emsimulin}. Subtracting this from the data leads to the distribution belonging to the 3BB channel which is also drawn in fig. \ref{Emsimulin}. 

\begin{figure}[t]
\begin{center}
\includegraphics[width=8cm,clip]{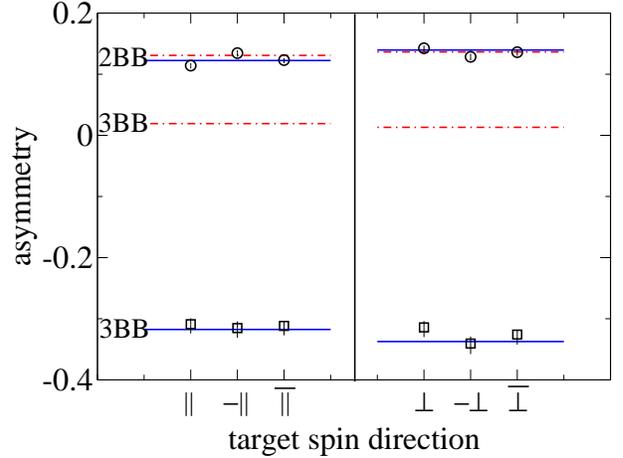}
\end{center}
\caption{\label{asym_all}Comparison of the data to the theoretical calculation for the 2BB and 3BB for the four target spin directions (anti)parallel ($\parallel$, --$\parallel$; left panel) and (anti)perpendicular ($\perp$, --$\perp$; right panel). In addition the combined sum for the parallel and perpendicular position is shown ($\overline{\parallel}$ and $\overline{\perp}$, respectively). To facilitate the comparison, all 2BB (3BB)data are shown with positive (negative) sign. PWIA: dot-dashed lines. Full calculation including FSI and MEC: solid lines. Statistical errors point up, systematic uncertainties point down. For the 2BB the size of the error bars is smaller than the symbols.}
\end{figure}

The ratio of the Monte Carlo simulation of the 2BB to the experimental data in the region of the 3BB is denoted by $a_{\rm 23}$. For the region 7.5 $<$ \emiss\ $<$ 25.5 MeV it amounts to  $a_{\rm 23}$ = 0.434 $\pm$ 0.002 (stat.) $\pm$ 0.015 (sys.). Then the asymmetry $A_{\rm 3BB}$ for the 3BB-channel is extracted from the asymmetry $A_{\rm 2+3BB}$ in the 3BB region (\emiss\ from 7.5 to 25.5 MeV) by accounting for the contribution from the radiation tail
\begin{equation} \label{corr_2BB}
A_{3BB} = \frac{A_{2+3BB} - A_{2BB} \, a_{23}}{1 - a_{23}} .
\end{equation}
All asymmetries are corrected for target and electron polarization. In fig. \ref{asym_all} the parallel and perpendicular asymmetries $A_{3BB}$ and $A_{2BB}$ are compared to two calculations of the Bochum-Krakow group. One uses PWIA only (dot-dashed), the other accounts for full FSI and MEC (solid line). The effect of MEC is negligible in this kinematics. The data integrated over the total detector acceptance are in good agreement with the calculation including FSI.    

The calculation shows that the FSI contribution is small in the 2BB
while it is large in 3BB. This suggests that the main contribution of FSI results from the rescattering term which does not exist in the 2BB, and not from direct FSI. This was also confirmed by further examination of the theoretical result by J. Golak \cite{Golak05}. 

In the 2BB channel the polarized \hepol\ can be interpreted as a polarized proton target. The spin of the neutron and the proton are parallel and couple to the deuteron spin equal to one. The spin of the second proton is aligned opposite to spin of the \hepol\ because the two protons reside mainly in S-state. In this configuration spin coupling and accounting for the Clebsch-Gordon coefficient revealed that a 100~\% polarized \hepol\ target corresponds to a 33~\% polarized proton in the 2BB channel. This is also confirmed by comparing the measured asymmetry $A_{2BB}$ with the one calculated for a free 100~\% polarized proton target.   

For the 3BB channel the situation is different. In PWIA the asymmetry is almost zero for the 3BB which reflects the fact that the two protons, which are dominantly in the $S$-state and thus have opposite spin orientation, now contribute equally to the knock-out reaction. The inclusion of FSI, however, leads to an asymmetry, which is larger and opposite in sign compared to the 2BB. The main effect comes from the np t-matrix (rescattering term). Since different spin combinations of the singlet and triplet np t-matrix contribute, the \hepol\ target cannot be interpreted as a polarized proton target in the 3BB channel. 

\section{Summary and Outlook} \label{sum}
In this contribution a review of the experiments with polarized \hepol\ performed at MAMI was given. The effort to build a machine to polarize \he\ started already in 1987. The first experiment at MAMI with \hepol\ was performed in the experimental hall A3 to measure the electric form factor of the neutron. Experiments with the same purpose at higher Q$^2$ followed, using improved target and detector setups in the three-spectrometer hall A1. With the new detector setup a better discrimination of inelastic events from the ones quasielastically scattered is possible. The performance of the target was steadily improved due to the development of new polarizers. This resulted in a more dense target (5 bar) with higher polarization ($\overline{P_T}$ = 50~\%). In addition the electron source was improved using a strained layer crystal. This led to nowadays available currents of 20~$\mu$A with an electron polarization of 75~\%. 

Parallel to the experiments the non-relativistic Faddeev calculation was developed by the Bochum-Krakow group. One of the first application was the calculation of the correction of \gen\ at Q$^2$ = 0.35~(GeV/c)$^2$ due to FSI which leads to a deviation of the measured asymmetry from that for a free neutron. This influence was also confirmed by measuring the target asymmetry A$_y$ where the beam is unpolarized and the target spin perpendicular to the scattering plane. This quantity is particularly sensitive to FSI and MEC contributions. In PWIA it vanishes. Good agreement between data and theory was found.

Another experiment concentrated on the question when a relativistic calculation is needed and which ingredients need to be treated relativistically. For this the reaction \heep\ at Q$^2$ = 0.67~(GeV/c)$^2$ was investigated. It turned out that the kinematics has to be treated relativistically already at this Q$^2$. On the other hand a relativistic current operator is much less important. At the moment a relativistic calculation is only possible in PWIA and with FSI23 included.    

To get more sensitive to the inner structure of \hepol\ the 2BB and 3BB channel in the reaction \heep\ were separated. Also here the theoretical calculation is in good agreement with the data. It is interesting that in the 2BB the \hepol\ target can be considered as a polarized proton target. This channel is almost not affected by FSI. However, large FSI effects are seen in the 3BB channel. 

All these reactions considered so far were not sensitive to MEC because the kinematics were chosen to correspond to the top of the quasielastic peak and the Q$^2$ was sufficiently high. At Q$^2$ $<$ 0.2~(GeV/c)$^2$ MEC contribute significantly to the reaction and modify the asymmetries. Since MEC's are not so well understood as compared to FSI it is planned to study kinematics which are sensitive to MEC. The data taken to measure \gen\ at Q$^2$ = 0.25~(GeV/c)$^2$ are affected by MEC in some kinematical regions covered by the detector acceptance.  

With the upgrade of MAMI to MAMI-C the \gen\ measurement will be pushed to Q$^2$ = 1.5~(GeV/c)$^2$. For this a new hadron detector is under construction which should have a higher neutron detection efficiency.

There are also plans to use polarized \hepol\ with (polarized) photons in the A2 experimental hall. Then \hepol\ would be used as a polarized neutron target to measure the Gerasimov-Drell-Hearn sum rule. For this a new target setup is needed which is already under consideration. 

\section*{Acknowledgments}
Finally I want to thank Mr. Kaiser for the excellent beam quality at MAMI and for his effort to adjust and setup the beam for our sensitive experiments. Then I want to thank J. Friedrich and Thomas Walcher for their support and advice as well as Mr. Arenh\"ovel, Mr. Backe and Mr. Drechsel for the good atmosphere in the institute.


\begin{thebibliography}{}
\bibitem{Golak02} J. Golak \et, Phys. Rev. \textbf{C65},  064004 (2002). 
\bibitem{Schulze93} R.W. Schulze and P.U. Sauer, Phys. Rev. \textbf{C48}, 38 (1993).
\bibitem{Blank84} B. Blankleider and R.M. Woloshyn, Phys. Rev. \textbf{C29}, 538 (1984).
\bibitem{Gao94} H. Gao \et, Phys. Rev. \textbf{C50}, R546 (1994).
\bibitem{Xu00} W. Xu \et, Phys. Rev. Lett. \textbf{85}, 2900 (2000).
\bibitem{Xu03} W. Xu \et, Phys. Rev. \textbf{C67}, 012201(R) (2003).
\bibitem{Meyerhoff94} M. Meyerhoff \et, Phys. Lett. B \textbf{327}, 201 (1994).
\bibitem{Becker99} J.~Becker \et, Eur.~Phys.~J. \textbf{A6}, 329 (1999).
\bibitem{Golak01} J. Golak \et, Phys. Rev. \textbf{C63}, 034006 (2001).
\bibitem{Rohe99} D.~Rohe \et, Phys. Rev. Lett. \textbf{83}, 4257 (1999).
\bibitem{Bermuth03} J.~Bermuth \et, Phys. Lett. B \textbf{564}, 199 (2003).
\bibitem{Achen05} P.~Achenbach \et, Eur. Phys. J. \textbf{A25}, 177 (2005).
\bibitem{Carasco03} C. Carasco \et, Phys. Lett. B {\bf 599}, 41 (2003).
\bibitem{Kamada02} H. Kamada, W. Gl\"ockle, J. Golak, Ch. Elster, Phys. Rev. {\bf C66}, 044010 (2002).
\bibitem{Walters62}G.K. Walters, F.D. Colgrove, and L.D. Schearer, Phys. Rev. Lett. \textbf{8}, 439 (1962).
\bibitem{Bouchiat60} M.A. Bouchiat, T.R. Carver, and C.M. Varnum, Phys. Rev. Lett. \textbf{5}, 373 (1960).
\bibitem{Haase98} D.G. Haase \et, Nucl. Instrum. \& Meth. \textbf{A402}, 341 (1998).
\bibitem{Goeck05} M. G\"ockeler \et, Nucl.Phys. {\bf A755}, 537 (2005).
\bibitem{Platch90} S. Platchkov \et, Nucl. Phys. {\bf A510}, 740 (1990).
\bibitem{Schia01} R. Schiavilla and I. Sick, Phys. Rev. {\bf C64}, 041002(R) (2001)
\bibitem{Arnold81} R.G. Arnold, C.E. Carlson, F. Gross, Phys. Rev. {\bf C23}, 363 (1981).
\bibitem{Herberg99}C.~Herberg \et, Eur. Phys. J. {\bf A5}, 131 (1999).
\bibitem{Ostrick99} M.~Ostrick \et, Phys. Rev. Lett. {\bf 83}, 276 (1999).
\bibitem{Thompson92} A.K. Thompson \et, Phys. Rev. Lett. {\bf 68}, 2901 (1992).
\bibitem{Eden94} T. Eden \et, Phys. Rev. {\bf C 50}, R1749 (1994).
\bibitem{Laget91} J.M. Laget, Phys. Lett. B {\bf 273}, 367 (1991).
\bibitem{Laget92} J.M. Laget, Phys. Lett. B {\bf 276}, 398 (1992).
\bibitem{Conzett98} H.E. Conzett, Nucl. Phys. {\bf A628}, 81 (1998).
\bibitem{Poolman99} H.R. Poolman, PhD. thesis, Vrije Universiteit te Amsterdam, 1999.
\bibitem{Wilms97} E.~Wilms \et, Nucl. Instr. \& Meth. {\bf A401}, 491 (1997).
\bibitem{Surkau97} R.~Surkau \et, Nucl. Instr. \& Meth. {\bf A384}, 444 (1997).
\bibitem{Passchier99}I.~Passchier \et, Phys. Rev. Lett. {\bf 82}, 4988 (1999).
\bibitem{Zhu01} H. Zhu \et, Phys. Rev. Lett. {\bf 87}, 081801 (2001).
\bibitem{Madey03} R. Madey \et, Phys. Rev. Lett. {\bf 91}, 122002 (2003).
\bibitem{Warren04} G. Warren \et, Phys. Rev. Lett. {\bf 92}, 042301 (2004).
\bibitem{Glazier05} D.I. Glazier \et, Eur.Phys.J. {\bf A24}, 101 (2005).
\bibitem{Galster71} S. Galster \et, Nucl. Phys. {\bf B32}, 221 (1971).
\bibitem{Fried03} J.~Friedrich and Th.~Walcher, Eur. Phys. J. A {\bf 17}, 607 (2003).
\bibitem{Ham04} H.-W. Hammer, D. Drechsel, Ulf-G. Mei{\ss}ner, Phys. Lett. B {\bf 586}, 291 (2004).
\bibitem{Machleidt96} R.~Machleidt, F.~Sammarruca, and Y.~Song, Phys. Rev., {\bf C53}, 1483 (1996).
\bibitem{Wiringa95} R.B. Wiringa, V.G.J. Stoks, and R.~Schiavilla, Phys. Rev., {\bf C51}, 38 (1995).
\bibitem{Deltuva04} A. Deltuva \et, Phys. Rev. {\bf C70}, 034004 (2004).
\bibitem{Aren92} H. Arenh\"ovel, W. Leidemann, E. Tomusiak, Phys. Rev. {\bf C46}, 455 (1992).
\bibitem{Otten04} E.W.~Otten, Europhysics News {\bf 35}, 16 (2004).
\bibitem{Flori99} R.E.J.~Florizone \et, Phys. Rev. Lett. {\bf 83}, 2308 (1999).
\bibitem{Golak05} J. Golak, private communication, 2005.
\end{thebibliography}
\end{document}